# Phase Retardation and Birefringence of the Crystalline Quartz Plate in the Ultraviolet and Visible Spectrum

**Xu Zhang** [1,2*], **Fuquan Wu** [2], **Limei Qi** [2], **Xia Zhang** [2], **Dianzhong Hao** [2]

*1. School of Physics Science and Engineering, Tongji University, Shanghai, 200092, China*

*2. Shandong Provincial Key Laboratory of Laser Polarization and Information Technology,
College of Physics and Engineering, Qufu Normal University, Qufu, 273165, China*

[*]*Corresponding author: zhangxuyq@163.com*

A method for measuring the phase retardation and birefringence of crystalline quartz wave plate in the ultraviolet and visible spectrum is demonstrated using spectroscopic ellipsometer. After the calibration of the crystalline quartz plate, the experimental data are collected by the photodetector and sent to the computer. According to the outputted data, the retardation can be obtained in the range of 190 to 770 nm. With the retardation data, the birefringence $(n_e - n_0)$ for the quartz can be calculated in the same spectrum with an accuracy of better than $10^{-5}$. The birefringence results enrich the crystalline quartz birefringence data especially in the ultraviolet spectrum.

## I. INTRODUCTION

Crystalline quartz, a natural birefringent crystal, is widely used in constructing wave plates. Quartz wave plate has important applications in laser technology[1,2], optical communications, etc. For a quartz wave plate, the phase retardation and birefringence need to be specified. At present, there are many methods to measure the phase retardation or birefringence. Based on normalized secondary harmonic component K. Yang gave a method for measuring the retardation of a quarter-wave plate[3]. Ying Zhang presented a method of wave plate phase retardation measurements using both direct laser lines and cross-wavelength measurement[4]. A. J. Zeng developed a method using one photoelastic modulator and one detector to measure the phase retardation and the fast axis angle of a quarter-wave plate simultaneously[5]. Polarization flipping method makes the phase retardation measurement feasible for online testing and has the advantage of auto-collimation [6-8]. The method in reference [9-12] can be used to gain the phase retardation or birefringence at one or several wavelengths, not the data in a spectral range.

In the application of wave plate, maybe we need to know the phase retardation of an arbitrary wave plate which might be a quarter-wave or a half-wave plate even a plate needed to be calibrated. So if a method can only be used to measure a kind of wave plate, its range of application will be very narrow. Recently, many applications require to measure the phase retardation at selected wavelengths. Life science instruments always require the accurate measurement of phase retardations at specific fluorescence wavelengths[4]. The method which can only be used to measure the phase retardation at one or several wavelengths can not meet the demand.

Reference [13-15] presented the birefringence of crystalline quartz at several wavelengths in the ultraviolet range. Since the birefringence of crystalline quartz is very sensitive to the variation of wavelength in the ultraviolet spectrum[13-15], there is great need to enrich the data about the birefringence of quartz in this spectrum. N.N. Nagib[16] presented the dispersion relations of the retardation and the birefringence for a multiple-order crystalline quartz plate when it introduces $\pi$ and $2\pi$ retardances by two optical arrangements. When the quartz plate introduces $\pi/2$ or $3\pi/2$ retardances, the data must be gained by interpolating results for $\pi$ and $2\pi$ retardances which will obviously decrease the accuracy of the birefringence. For a multiple-order quartz plate, a small error in the value of birefringence will introduce a corresponding serious error in the retardation value. Furthermore, the birefringence at the wavelengths except mentioned above can not be gained.

In this work, we develop a method to measure the phase retardation of a crystalline quartz plate between 190 and 770 nm using the spectroscopic ellipsometer. With the phase retardation data we can gain the birefringence of the crystalline quartz in the same spectrum. We don't need to know the type of the wave plate and its fast axis in advance. To demonstrate this powerful technique, we use a crystalline quartz plate of thickness 1.625 mm as an example to illustrate the measurement principles. Excellent results are obtained.

## II. MEASURING PRINCIPLE

The standard ellipsometry mainly includes the reflection ellipsometry and the transmission ellipsometry. As we all know, ellipsometric angles $\psi$ and $\Delta$ are two important parameters in the ellipsometry method. In the reflection ellipsometry, as to the isotropic medium the relationship between the ellipsometry parameters and the reflection matrix $\begin{bmatrix} r_p & 0 \\ 0 & r_s \end{bmatrix}$ is [17]:

$$\rho = \frac{r_p}{r_s} = \tan\psi e^{i\Delta} . \quad (1)$$

But to the anisotropic medium the reflection matrix is not diagonalized, it is $\begin{bmatrix} r_{pp} & r_{sp} \\ r_{ps} & r_{ss} \end{bmatrix}$. The relationship between the ellipsometry parameters and this matrix is [18]:

$$\begin{cases} \rho_{ps} = \dfrac{r_{ps}}{r_{pp}} = \tan\psi_{ps}\exp(i\Delta_{ps}) \\ \rho_{sp} = \dfrac{r_{sp}}{r_{ss}} = \tan\psi_{sp}\exp(i\Delta_{sp}) \\ \rho = \dfrac{r_{pp}}{r_{ss}} = \tan\psi\exp(i\Delta) \end{cases} \quad (2)$$

The reflection matrix can be diagonalized as $\begin{bmatrix} r_p & 0 \\ 0 & r_s \end{bmatrix}$ when the optic axis is parallel or vertical to the incident plane with

$$r_p = f_1(n_o, n, d), \quad r_s = f_2(n'_e, n, d). \quad (3)$$

Where $n$ is the refractive index of the background medium, $n_o$ is the refractive index of $o$ light ($s$ vibration) in the plate, $d$ is the thickness of the plate. And $n'_e = \dfrac{n_o n_e}{\sqrt{n_e^2 \cos^2\theta + n_o^2 \sin^2\theta}}$ is the refractive index of $e$ light ($p$ vibration) in the plate (shown in Figure 1). It is evident that (3) is also very complicated, we need complex data processing to find the phase retardation and birefringence of a wave plate.

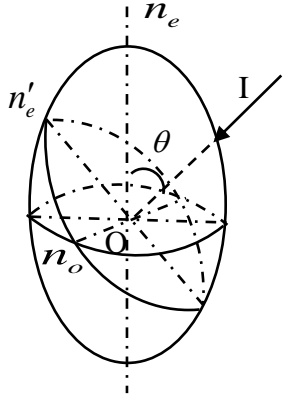

Figure 1 Indicatrix of crystalline quartz.

As to the standard transmission ellipsometry, suppose the complex amplitude transmission coefficient of the $p$ vibration in the transmission wave is $t_p$, then $t_p = |t_p|\exp(i\Delta_{tp})$ ($\Delta_{tp}$ is the phase shift of the $p$ vibration). And the complex amplitude transmission coefficient of the $s$ vibration is $t_s = |t_s|\exp(i\Delta_{ts})$ ($\Delta_{ts}$ is the phase shift of the $s$ vibration). The ratio of $t_p$ to $t_s$ which can be measured by the ellipsometer is shown as

$$\rho_t = \dfrac{t_p}{t_s} = \tan\psi_t \exp(i\Delta_t). \quad (4)$$

Where

$$\psi_t = \tan^{-1}|\rho_t|, \quad (5)$$

$$\Delta_t = \Delta_{tp} - \Delta_{ts} \ . \qquad (6)$$

When a linearly polarized light vertically entrances a crystalline quartz wave plate, it is decomposed into $o$ light ($s$ vibration) and $e$ light ($p$ vibration). The refractive index is $n_o$ and $n_e$ respectively. Because the two lights have different speeds in the wave plate, after they passed through the wave plate with the thickness of $d$ the absolute phase retardation $\Delta$ between them is [16]:

$$\Delta = 2\pi(n_e - n_0)d/\lambda = 2\pi m + \varphi \ , \qquad (7)$$

where $(n_e - n_0)$ is the birefringence at the wavelength $\lambda$ and $\varphi$ is the apparent retardation ($0 \le \varphi < 2\pi$). In general quartz wave plates are manufactured as multiple-order plates. So $m$ is a positive integer (the order of the plate at $\lambda$).

From the above discussion, the transmission ellipsometry can be used to measure the phase retardation of the quartz wave plate. The UVISEL spectroscopic phase modulated ellipsometer (shown in Figure 2) made by the French Jobin Yvon corporation is used in the experiment. $L$ is laser, $P$ the polarizer, $S$ the measured wave plate, $A$ and $A'$ the analyzers, $M$ the modulator, $M'$ the monochromator, $D$ the detector and $C$ the computer. It has a continuous spectral range of 190-1700nm. The accuracy of the angular instrument is $0.05°$ which is also the accuracy of measuring other angles in the optical system.

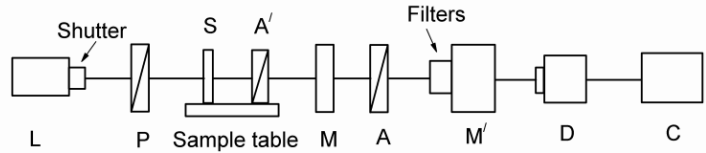

Figure 2 Optical system for measuring the phase retardation

and birefringence of crystalline quartz plate.

As to the UVISEL spectroscopic phase modulated ellipsometer, the polarizer $P$, modulator $M$ and analyzer $A$ have two main configurations. Configuration I is $P - M = 45°$, $M = 0°$, $A = 45°$ and configuration II is $P - M = 45°$, $M = 45°$, $A = 45°$. In configuration I, the final detected intensity collected by the detector is

$$I(t) = I[I_0 + I_s \sin\delta(t) + I_c \cos\delta(t)] \ . \qquad (8)$$

With

$$I_0 = 1, \quad I_s = \sin 2\psi \sin\Delta, \quad I_c = \sin 2\psi \cos\Delta . \qquad (9)$$

While in configuration II, the final detected intensity collected by the detector is

$$I(t) = I[I_0 + I_s \sin\delta(t) + I_c \cos\delta(t)] . \qquad (10)$$

With

$$I_0 = 1, \quad I_s = \sin 2\psi \sin\Delta, \quad I_c = \cos 2\psi . \qquad (11)$$

From the above equations we can see that in configuration I we can measure $\Delta$ accurately over the full $0° - 360°$ range (theoretical equations (9) give $\tan\Delta$ and $\sin 2\psi$). But suffer from $\psi$ accuracy problems when $\psi$ is near $45°$. On the contrary in the configuration II, we can measure $\psi$ accurately over the full $0° - 90°$ range, but suffer from $\Delta$ indetermination in the range $90° - 270°$. Since we want to gain the phase retardation $\Delta$ of the quartz plate, we use the configuration I in the experiment.

In the measurement the optic axis should be vertical to the incident light. If the angle between the optic axis and the incident light is $\theta$, what we gain is $n'_e - n_0$ ($n'_e = \frac{n_o n_e}{\sqrt{n_e^2 \cos^2 \theta + n_o^2 \sin^2 \theta}}$) which can be seen from Figure 1. When $\theta = \frac{\pi}{2}$, $n'_e = n_e$ and $n'_e - n_0 = n_e - n_0$.

## III. EXPERIMENT

The experiment procedures are as follows:

1) Polarizer $P$, modulator $M$ and analyzer $A$ are aligned in orientations with $P = 0°$, $M = 0°$, $A = 45°$. Then place another analyzer $A'$ on the sample table. Adjust $A'$ to make $P$ and $A'$ crossed.

2) Then place the measured plate between $P$ and $A'$, adjust the plate until minimum intensity (within the measurement noise level) is detected which means the optic axis of the measured plate is parallel or vertical to the transmission axis of the polarizer $P$. In the mean time the first harmonic component of the modulated intensity $R\omega$ and the second harmonic component $R2\omega$ should be zero (within the measurement noise level) which enables us to use the conventional ellipsometric angles $\psi$ and $\Delta$ in the standard ellipsometric configurations to measure an uniaxially anisotropic slab (the quartz wave plate) [19].

3) Set $P = 90°$, then adjust $A'$ to make $P$ and $A'$ crossed again. If the first and the second harmonic components of the modulated intensity are still zero. Then the optic axis of the measured plate is vertical to the incident light [19]. In our experiment $R\omega = 0.00048$, $R2\omega = 0.00031$ in the two configurations.

4) Take $A'$ away and adjust $P = 45°$, set the spectral range and wavelength interval of the UVISEL ellipsometer, the retardation of the wave plate can be gained from the calculation of the outputted data.

5) Then adjust $P = -45°$ to repeat the measurement in procedure 4).

In this experiment the spectral range is $190 - 770$ nm and the wavelength interval is 10 nm. The interval can be set as 5 nm, 1 nm and so on. So we can gain more data when the interval is smaller. The data of interval 10 nm are presented in Table 1 for simplicity. The procedure 5) is used to check the p and s transmissions are same or not. If p and s transmissions are different, there must be miscuts in the preparation of the wave plate. Then analysis schemes must invoke the full anisotropic generalized ellipsometry algorithm[20]. In our experiment the p and s transmissions agree well with each other.

## IV. RESULTS and DISCUSSIONS

The experiment is made at the temperature $25 °C \pm 0.1 °C$. The thickness of the plate is measured by a digital micrometer with an accuracy of $0.5 \mu m$ at different points to decrease the influence of thickness variation. The mean value of the thickness $d$ is 1.625 mm. From reference [13] we recall that $(n_e - n_0) = 0.0098$ at 350nm for crystalline quartz. We substitute this value at 350 nm. With $d = 1.625$ mm, it follows from equation (7) that $\Delta = 16380° = 90\pi + 160°$. It is concluded that the order of the plate at this wavelength is $m = 45$. As to our experimental data, from the second column in Table 1 we can know that the apparent retardation $\varphi = 146.091°$ at 350nm. So the absolute phase retardation should be $\Delta = 90\pi + 146.091° = 16346.091°$. Then from equation (7) we can know that the birefringence for the quartz plate at 350nm is $(n_e - n_0) = 0.00978$ which agrees well with reference [13]. With the same method the birefringence at other wavelengths can be gained.

The apparent phase retardation $\delta$, absolute phase retardation $\Delta$ and birefringence for crystalline quartz plate are listed in Table 1. From the apparent phase retardation data this quartz plate with a thickness of 1.625 mm is quarter-wave plate at 270 nm and 470nm. It is a half-wave plate at 220nm. With a smaller wavelength interval, we can gain more data.

Table 1 Experimental apparent retardation $\delta$, absolute phase retardation $\Delta$, and birefringence for crystalline quartz plate at $25\ °C \pm 0.2\ °C$

| $\lambda$ (nm) | $\varphi$ /(deg) | $\Delta$ /(deg) | $(n_e - n_o) \times 10^{-3}$ | Data[13] $\times 10^{-3}$ |
|---|---|---|---|---|
| 190 | 5.125 | 42485.125 | 13.799 | |
| 200 | 196.103 | 37996.103 | 12.990 | 13.0 |
| 210 | 200.657 | 34580.657 | 12.414 | |
| 220 | 185.297 | 31865.297 | 11.984 | |
| 230 | 255.816 | 29595.816 | 11.636 | |
| 240 | 275.093 | 27635.093 | 11.337 | |
| 250 | 60.904 | 25980.904 | 11.103 | 11.1 |
| 260 | 191.078 | 24491.078 | 10.885 | |
| 270 | 90.814 | 23130.814 | 10.676 | |
| 280 | 166.721 | 21946.721 | 10.504 | |
| 290 | 74.523 | 20954.523 | 10.388 | |
| 300 | 112.009 | 20092.009 | 10.304 | 10.3 |
| 310 | 311.849 | 19211.849 | 10.181 | |
| 320 | 352.620 | 18352.620 | 10.039 | |
| 330 | 307.897 | 17587.897 | 9.921 | |
| 340 | 113.333 | 17033.333 | 9.899 | |
| 350 | 146.091 | 16346.091 | 9.780 | 9.80 |
| 360 | 317.606 | 15797.606 | 9.722 | |
| 370 | 139.399 | 15259.399 | 9.651 | |
| 380 | 240.802 | 14820.802 | 9.627 | |
| 390 | 350.738 | 14390.738 | 9.594 | |
| 400 | 288.930 | 13968.930 | 9.551 | 9.60 |
| 410 | 259.712 | 13579.712 | 9.517 | |
| 420 | 109.926 | 13069.926 | 9.383 | |
| 430 | 105.055 | 12705.055 | 9.339 | |
| 470 | 86.121 | 11606.121 | 9.325 | |
| 490 | 285.590 | 11085.590 | 9.285 | |
| 510 | 195.518 | 10635.518 | 9.272 | |
| 530 | 76.814 | 10156.814 | 9.202 | |
| 550 | 33.829 | 9753.829 | 9.170 | 9.17 |
| 570 | 20.465 | 9380.465 | 9.140 | |
| 590 | 37.187 | 9037.187 | 9.114 | |
| 620 | 284.033 | 8564.033 | 9.076 | |
| 630 | 222.781 | 8502.781 | 9.062 | |
| 650 | 221.782 | 8141.782 | 9.046 | 9.03 |
| 670 | 319.836 | 7879.836 | 9.024 | |
| 680 | 194.011 | 7754.011 | 9.013 | |
| 700 | 40.517 | 7600.517 | 8.995 | 8.98 |
| 730 | 347.036 | 7187.036 | 8.968 | |
| 740 | 247.374 | 7087.374 | 8.965 | |
| 770 | 314.232 | 6794.232 | 8.943 | |

The absolute phase retardation of the crystalline quartz plate is shown in Figure 3 from Table 1. In Figure 4(a) our birefringence data of crystalline quartz are represented by the black dots. Expressing birefringence of the crystalline quartz as a power series of the form (the unit of $\lambda$ is nm):

$$(n_e - n_0) = 0.29765787 - 5.58911 \times 10^{-3}\lambda + 4.828 \times 10^{-5}\lambda^2 - 2.40755 \times 10^{-7}\lambda^3$$
$$+ 7.59404 \times 10^{-10}\lambda^4 - 1.56865 \times 10^{-12}\lambda^5 + 2.12158 \times 10^{-15}\lambda^6$$
$$- 1.8127 \times 10^{-18}\lambda^7 + 8.88594 \times 10^{-22}\lambda^8 - 1.90596 \times 10^{-25}\lambda^9. \quad (12)$$

We neglect the higher order terms and the fitting red curve is shown in Figure 4(b).

Data of reference [14] are also widely used[16] represented by the blue stars in Figure 4(c). In Figure 4(d) the data of reference [15] are represented by the red stars. From the two figures, we can see that our birefringence data highly agree with the previously published data[14,15] for crystalline quartz. What's more, our data are richer especially in the ultraviolet spectrum.

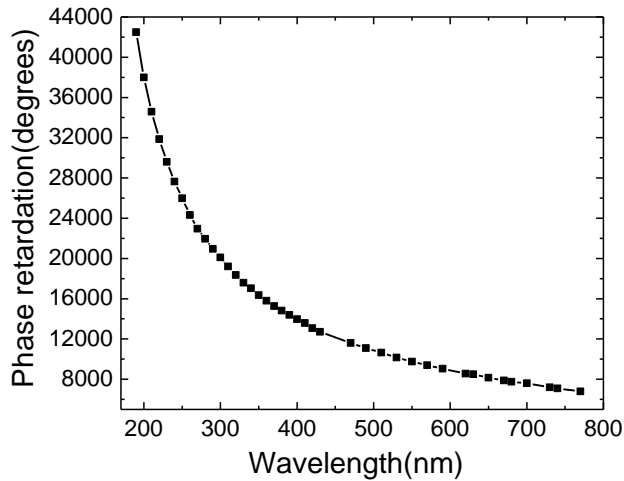

Figure 3 Absolute phase retardation curve for the crystalline quartz plate in Table 1.

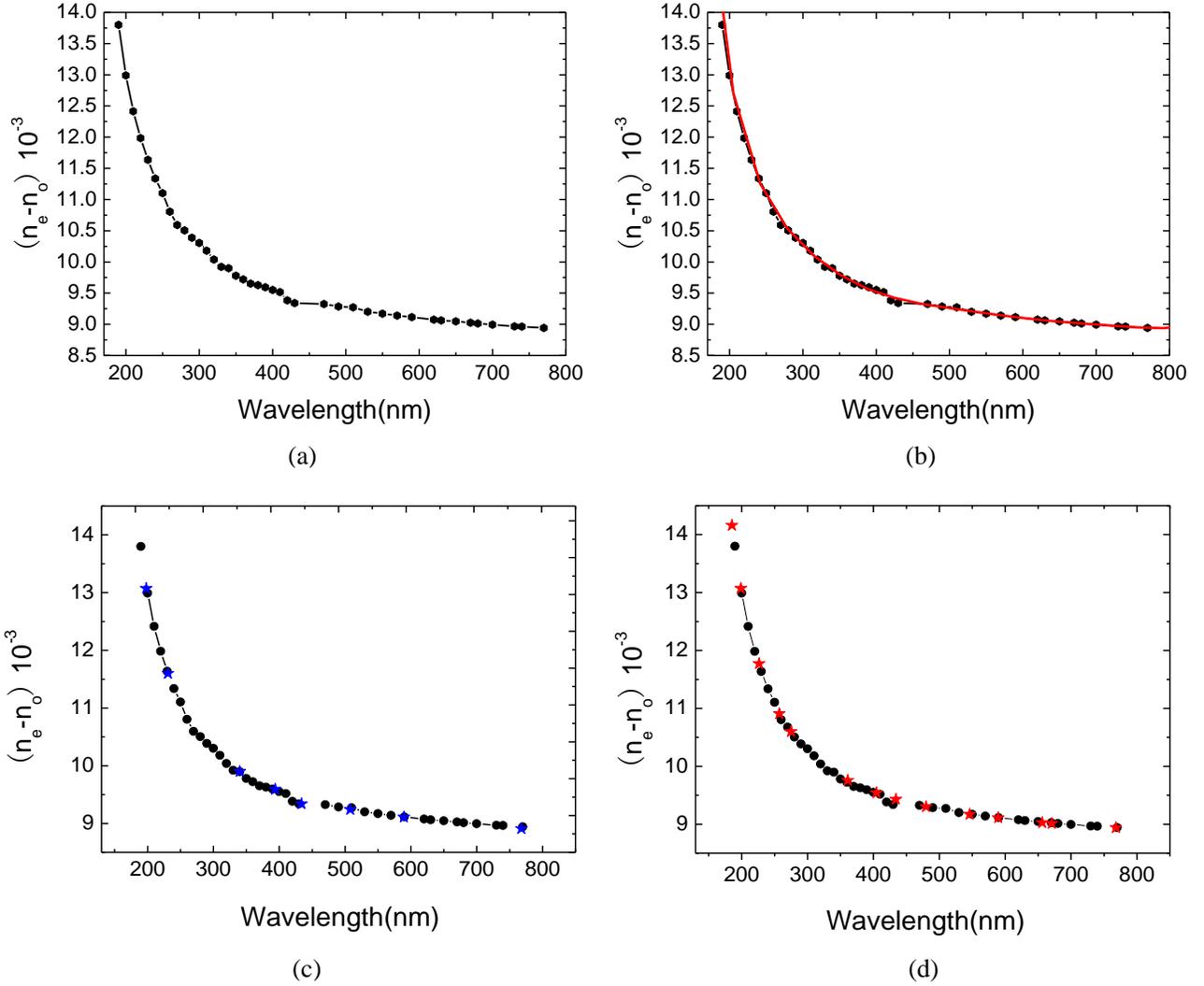

Figure 4 (a) Birefringence data for the crystalline quartz represented by the black dots from Table 1. (b) (Color online) The polynomial fit curve of our birefringence data represented by the red curve. (c) (Color online) Data of reference [14] represented by the blue stars. (d) (Color online) Data of Gorachand Ghosh in reference [15] represented by the red stars.

From equation (7) we can know $(n_e - n_0) = \dfrac{\lambda \Delta}{2\pi d}$, so the maximum error of $(n_e - n_0)$ is

$$\delta(n_e - n_0) = \dfrac{\delta \lambda}{2\pi d} \Delta + \dfrac{\delta \Delta}{2\pi d} \lambda + \dfrac{\lambda \Delta}{2\pi d^2} \delta d \qquad (13)$$

As to the UVISEL spectroscopic phase modulated ellipsometer, the accuracy of the wavelength is 0.1 nm. So the maximum value of $\dfrac{\delta \lambda}{2\pi d} \Delta$ is $7.3 \times 10^{-6}$ (the maximum value of $\Delta$ is 42485.125° at 190nm).

Through repeated measurement, we know that the maximum error of $\Delta$ is $\pm 0.3°$, so that the maximum value of $\dfrac{\delta\Delta}{2\pi d}\lambda$ is $3.9\times10^{-7}$ (when $\lambda$ is 770nm) in the $190-770$ nm spectrum.

Since the digital micrometer has an accuracy of $0.5\mu m$, the third part in equation (13) $\dfrac{\lambda\Delta}{2\pi d^2}\delta d$ has a maximum value:

$$\dfrac{\lambda\Delta}{2\pi d^2}\delta d = \dfrac{190\times10^{-9}\times 42485.125}{360\times(1.625\times10^{-3})^2}\times 0.5\times10^{-6} = 4.2\times10^{-6}$$ (from Table 1 we find that $\lambda\Delta$ gains the maximum value at 190nm).

The accuracy of the angular instrument in the UVISEL ellipsometer is $0.05°$ which determines the birefringence value is $(n_e - n_0)$ or $(n_e' - n_0)$. Through calculation we know the maximum value of $n_e - n_e' = n_e - \dfrac{n_o n_e}{\sqrt{n_e^2\cos^2\theta + n_o^2\sin^2\theta}}$ is less than $1\times10^{-7}$.

The accuracy of the thermometer is $0.1\ °C$ in the experiment. From another experiment (not shown here) aimed to find the temperature effect of quartz birefringence we find that the maximum change of the $(n_e - n_0)$ is $2\times10^{-6}$ in the $190-770$ nm spectrum when the temperature changes $1\ °C$. So the error of $(n_e - n_0)$ is $2\times10^{-7}$ arises from the temperature accuracy in our experiment.

From the above error analysis, the accuracy of birefringence $(n_e - n_0)$ is better than $10^{-5}$ for most wavelengths in the ultraviolet spectrum, while in the visible spectrum better than $10^{-6}$. The accuracy of $(n_e - n_0)$ in the visible spectrum is better than the ultraviolet spectrum mainly because the accuracy of the whole system has a greater influence to the short wave than the long wave.

## V. CONCLUSION

We have demonstrated a method to measure the phase retardation and birefringence of crystalline quartz plate using a spectroscopic ellipsometer. From error analysis the data of quartz birefringence has an accuracy of better than $10^{-5}$ in the $190-770$ nm spectrum. The method described above has the following advantages. The first is that the phase retardation is obtained in a large spectral range rather than at one or several wavelengths. And we will know the wave plate can act as a quarter-wave or a half-wave plate at many wavelengths which will lead to the full use of it. Secondly, the birefringence of the quartz wave plate can be calculated from the retardation data at the wavelengths in the same spectral range. Finally our birefringence results agree well with the previously published data and enrich the quartz birefringence data especially in the ultraviolet spectrum.

## ACKNOWLEDGMENT

This study was supported by the National Youth Natural Science Foundation of China (No. 61107030).

## REFERENCES

1. Z.H. Hu, S. L. Zhang, W. X. Liu and H. B. Jia, "The correlation between frequency difference and mode polarization in an intracavity birefringent dual-frequency laser," Opt. Laser Technol., **42**, 540-545(2010).
2. Liang Miao, Duluo Zuo, Zhixian Jiu and Zuhai Cheng, "An efficient cavity for optically pumped terahertz lasers, "Opt. Commun. , **283**, 3171-3175(2010).
3. K. Yang, A. J. Zeng, X. Z. Wang and H. Wang, "Method for measuring retardation of a quarter-wave plate based on normalized secondary harmonic component," Optik, **120**, 558-562(2009).
4. Y. Zhang, F. Song, H. Li, and X. Yang, "Precise measurement of optical phase retardation of a wave plate using modulatedpolarized light," Appl. Opt. 49, 5837–5843 (2010).
5. A. J. Zeng, F.Y. Li, L. L. Zhu, and H. J. Huang, "Simultaneous measurement of retardance and fast axis angle of a quarter-wave plate using one photoelastic modulator, " Appl. Opt., 50, 4347-4352(2011).
6. W. X. Chen, S. L. Zhang, X. W. Long, Locking phenomenon of polarization flipping in He-Ne laser with a phase anisotropy feedback cavity. Appl Opt, **51**, 888–893(2012).
7. W. X. Chen, S. L. Zhang, X. W. Long, et al. Phase retardation measurement by analyzing flipping points of polarization states in laser with an anisotropy feedback cavity. Opt Laser Technol, **44**, 2427–2431(2012).
8. W. X. Chen, H. H. Li, S. L. Zhang, et al. Measurement of phase retardation of wave plate online based on laser feedback. Rev Sci Instrum, **83**, 013101–013101–3 (2012).
9. T. C. Chey; S. M. Lee, Determination of Birefringence by Brillouin Spectroscopy, Journal of the Optical Society of Korea, **2**, 45-49 (1998).
10. S. M. Lee; C. K. Hwangbo, Measurement of Birefringence with Brillouin Spectroscopy Journal of the Optical Society of Korea, **5**, 67-69 (2001).
11. Y. Zhang, S. L. Zhang, Y.M. Han, Y. Li and X. N. Xu, "Method for the measurement of retardation of wave plates based on laser frequency-splitting technology," Opt. Eng. **40**, 1071-1075(2001).

12. X. Zhang, F.Q. Wu, H. L. Wang, B. Yan and C. Kong, "Simultaneous measurement of the phase retardation and optic axis of wave plates," Optoelectronics Letters, **3**, 65-68(2007).
13. Jean M. Bennett and Harold E. Bennett, "Polarization," (In: Driscoll WG, editor. Handbook of optics. New York: McGraw-Hill) (1978).
14. William L. Wolfe, "Properties of Optical Materials," (In: Driscoll WG, editor. Handbook of optics. New York: McGraw-Hill) (1978).
15. Gorachand Ghosh, "Dispersion-equation coefficients for the refractive index and birefringence of calcite and quartz crystals," Opt. Commun. **163**, 95-102(1999).
16. N.N. Nagib, S.A. Khodier and H.M. Sidki, "Retardation characteristics and birefringence of a multiple-order crystalline quartz plate," Opt. Laser Technol. **35**, 99-103(2003).
17. A. Laskarakis, S. Logothetidis, E. Pavlopoulou, and M. Gioti, "Mueller matrix spectroscopic ellipsometry: formulation and application," Thin Solid Films, **455-456**, 43-49(2004).
18. Volodymyr Tkachenko, Antigone Marino, Francesco Vita, D'Amore F, De Stefano L, Malinconico M, Rippa M and Abbate G, "Spectroscopic ellipsometry study of liquid crystal and polymeric thin films in visible and near infrared," Eur. Phys. J. E **14**, 185-192(2004).
19. H. Touir, M. Stchakovsky, R. Ossikovski and M. Warenghem, "Coherent and incoherent interference modelling and measurement of anisotropic multilayer stacks using conventional ellipsometry," Thin Solid Films, **455-456**, 628-631(2004).
20. M. Schubert, "Another century of ellipsometry", Annalen der Physik 15, 480-497, (2006).